\documentclass[twocolumn,showpacs,pre]{revtex4}
\usepackage{dcolumn}
\usepackage{bm}

\begin{document}

\title{Towards a Quantum Fluid Mechanical Theory of Turbulence}

\author{D. Drosdoff, A. Widom, J. Swain}
\affiliation{Physics Department, Northeastern University, Boston MA USA}
\author{Y.N. Srivastava}
\affiliation{Physics Department \& INFN,  University of Perugia, Perugia IT}
\author{ V. Parihar}
\affiliation{Physics Department, Boston University, Boston MA USA}
\author{S. Sivasubramanian}
\affiliation{NSF NSEC Center for High-rate Nanomanufacturing \\ 
Northeastern University, Boston MA USA}

\begin{abstract}
Recent studies of turbulence in superfluid Helium indicate that  
turbulence in quantum fluids obeys a Kolmogorov scaling law. 
Such a law was previously attributed to classical solutions of the 
Navier-Stokes equations of motion. It is suggested that turbulence in 
all fluids is due to quantum fluid mechanical effects. 
Employing a field theoretical view of the fluid flow velocity, vorticity 
appears as quantum filamentary strings. This in turn leads directly to the 
Kolmogorov critical indices for the case of fully developed turbulence.    
\end{abstract}

\maketitle

\section{Introduction \label{INTRO}}

It is often thought that fluid mechanical turbulence is the last of the 
great unsolved problems of classical physics\cite{LandauLifshitz:1987}. 
Even with extensive numerical solutions of the classical Navier-Stokes 
fluid mechanical equations of motion, it has not been possible to prove 
simple experimental rules, which have been formulated for turbulent fluid 
flows\cite{McComb:1990}. Our purpose is to suggest that turbulence is 
inherently a quantum mechanical problem which cannot be solved by 
classical methods.

To see what is involved in a quantum mechanical picture, let us consider 
the role of quantum mechanics in describing vorticity and 
turbulence. The classical Reynolds number describing 
a turbulent states is defined as 
\begin{equation}
{\cal R}=\frac{\cal VL}{\nu}=\frac{\cal VL\rho }{\eta }\ ,
\label{INTRO1}
\end{equation}
wherein \begin{math}  {\cal V}  \end{math} and \begin{math} {\cal L}  \end{math} 
represent, respectively, the velocity and length scales of the flow, 
\begin{math}  \nu  \end{math} and \begin{math}  \eta  \end{math} represent, 
respectively, kinematic and dynamic viscosity of the fluid and 
\begin{math}  \rho  \end{math} is the fluid mass density. Turbulent flows 
are normally described by high Reynolds number. The number of  
quantum vortex filaments characterizing a flow as described by 
Feynman\cite{Feynman:1955} reads 
\begin{equation}
{\cal R}_F=\frac{\cal VL}{\kappa },
\label{INTRO2}
\end{equation}
wherein the quantized vortex circulation \begin{math} \kappa \end{math} for a 
fluid containing molecules each of mass \begin{math} M \end{math} and total 
atomic mass number 
\begin{math} A \end{math} is given by 
\begin{equation}
\kappa=\frac{2\pi \hbar}{M}\approx 
\frac{3.96\times 10^{-3}}{A}\ \frac{\rm cm^2}{\rm sec}\ .
\label{INTRO3}
\end{equation}  
For the formation of a single quantum vortex in an otherwise 
circulation free fluid
flow, the Feynman number \begin{math} R_F\sim 1 \end{math}. The ratio 
\begin{equation}
\frac{{\cal R}_F}{\cal R}=\frac{\nu }{\kappa }   
\label{INTRO4}
\end{equation}
denotes the number of quantized filaments within a turbulent tube normally 
described employing {\em classical} vorticity. For water 
\begin{math} {\rm H_2O} \end{math}\cite{Widom:1990} with \begin{math} A=18  \end{math} 
and at room temperature and pressure, 
\begin{equation}
\left[\frac{{\cal R}_F}{\cal R}\right]_{water}=\frac{\nu_{water}}{\kappa_{water}}
\approx 45. 
\label{INTRO5}
\end{equation}
Vortex tubes in turbulent water flows may then be safely considered to be constructed 
from quantized vortices.
 
In Sec.\ref{QFM} the quantized theory of fluid mechanics will be reviewed\cite{Landau:1941} 
and formulated as a Lagrangian quantum field theory. Quantized vortex filaments will then 
be discussed in Sec.\ref{VF}. In Sec.\ref{CRQ} the general theory of quantum fluid 
quantization will be written in the Clebsch representation\cite{Clebsch:1859}. 
In Sec.\ref{KC} the statistical correlation functions of velocity and vorticity are 
defined and discussed.   The classical Kolmogorov 
model\cite{Kolmogorov:1941a,Kolmogorov:1941b,Kolmogorov:1941c} 
for fully developed turbulence will then be considered wherein the critical 
index for velocity fluctuations will be derived. In Sec.\ref{FDT}, the Kolmogorov scaling
exponents will be derived under the assumption that quantized vortex filaments have the 
geometry characteristic of a self avoiding random walk. When a path is embedded in 
\begin{math} D \end{math} spatial dimensions, the fractal dimension of a random walk path  
is two, while the fractal dimension of a {\em self-avoiding} random walk path is 
\begin{eqnarray}
d=\frac{D+2}{3},
\nonumber \\ 
d=5/3\ \ \ {\rm for}\ \ \ D=3.
\label{Intro6}
\end{eqnarray} 
The fractal dimension \begin{math} d=5/3  \end{math} of a random 
quantized vortex line uniquely determines the Kolmogorov scaling exponents 
for fully developed turbulent quantum fluid dynamics. In the concluding 
Sec.\ref{CONC}, the very similar nature of turbulence in the quantum superfluid phase of 
\begin{math} ^4He_2   \end{math} and \begin{math} ^3He-B \end{math} and in ordinary 
fluids usually regarded as classical 
will be discussed. It will be argued that turbulence in fluid flows {\em normally regarded as classical} 
are in reality inherently quantum mechanical in nature. 

\section{Quantum Fluid Mechanics \label{QFM}}

Let \begin{math} M_a  \end{math} be the mass of the \begin{math} a^{th} \end{math} molecule 
at position \begin{math} {\bf r}_a  \end{math} in a fluid. The mass density 
\begin{math} \rho  \end{math} and the mass current  
\begin{math} {\bf J}  \end{math} operators may be defined  
\begin{eqnarray}
\rho ({\bf r})&=&\sum_a M_a\delta ({\bf r}-{\bf r}_a),
\nonumber \\ 
{\bf J}({\bf r})&=&\frac{1}{2}\sum_a 
\big({\bf p}_a\delta ({\bf r}-{\bf r}_a)+\delta ({\bf r}-{\bf r}_a){\bf p}_a\big),
\label{QFM1}
\end{eqnarray}
wherein the momentum \begin{math} {\bf p}_a  \end{math} conjugate to the position 
\begin{math} {\bf r}_a  \end{math} obeys 
\begin{equation}
\left[{\bf p}_a ,{\bf r}_b \right]=-i\hbar {\bf 1}\delta_{ab}.
\label{QFM2}
\end{equation}
The mass density and the mass current thereby obey the equal time commutation relations 
\begin{equation}
\frac{i}{\hbar }\left[{\bf J}({\bf r}) ,\rho({\bf r}^\prime ) \right]
=\rho ({\bf r}){\bf grad}\ \delta ({\bf r}-{\bf r}^\prime).
\label{QFM3}
\end{equation}
One may formally define the fluid velocity operator \begin{math} {\bf v}  \end{math} 
according to 
\begin{equation}
{\bf J}({\bf r})=\frac{1}{2}\big(\rho ({\bf r}){\bf v}({\bf r})
+{\bf v}({\bf r})\rho ({\bf r})\big)
\label{QFM4}
\end{equation}
in which case the velocity field does not commute with the mass density field; i.e. 
\begin{equation}
\frac{i}{\hbar }\left[{\bf v}({\bf r}) ,\rho({\bf r}^\prime ) \right]
={\bf grad}\ \delta ({\bf r}-{\bf r}^\prime).
\label{QFM5}
\end{equation}
Note the consequences of the Jacobi identities which follow from Eq.(\ref{QFM5}) 
in the form 
\begin{eqnarray}
\left[v_i({\bf r}),\left[v_j({\bf r}^\prime ),\rho({\bf r}^{\prime \prime }) \right]\right]
&=& 0,
\nonumber \\
\left[\rho({\bf r}^{\prime \prime }) ,\left[v_i({\bf r}^\prime ),v_i({\bf r})\right]\right]
&=& 0.
\end{eqnarray}
The mass density commutes with the velocity component commutators. 
By taking the curl of both sides of Eq.(\ref{QFM5}). it follows that the vorticity 
components also commute with the mas density,
\begin{eqnarray}
{\bf \Omega}({\bf r})={\rm curl}{\bf v}({\bf r}),
\nonumber \\ 
\left[{\bf \Omega }({\bf r}) ,\rho({\bf r}^\prime ) \right]=0.
\label{QFM7}
\end{eqnarray}
The commutation relations between different components of the velocity read
\begin{eqnarray}
\frac{i}{\hbar }\rho ({\bf r})\left[v_i({\bf r}),v_j({\bf r}^\prime ) \right]
=\frac{i}{\hbar }\left[v_i({\bf r}),v_j({\bf r}^\prime ) \right]\rho ({\bf r}^\prime )
\nonumber \\ 
\frac{i}{\hbar }\rho ({\bf r})\left[v_i({\bf r}),v_j({\bf r}^\prime ) \right]
=\Omega_{ij}({\bf r})\delta ({\bf r}-{\bf r}^\prime ),
\nonumber \\ 
\Omega_{ij}({\bf r})=\epsilon_{ijk}\Omega_k({\bf r}).
\label{QFM8}
\end{eqnarray} 
Let us now consider the quantum vorticity and fluid circulation in more detail.

\section{Vortex Filaments \label{VF}}

For a given velocity potential function \begin{math} \Phi ({\bf r}) \end{math}, 
consider the unitary operator 
\begin{equation}
U=\exp \left(\frac{i}{\hbar}\int \rho({\bf r})\Phi ({\bf r})d^3{\bf r}\right).
\label{VF1}
\end{equation}
Acting on a wave function \begin{math} \Psi  \end{math} 
of \begin{math} N \end{math} fluid molecules, 
\begin{equation}
U\Psi =\exp \left(\frac{iM}{\hbar}\sum_{a=1}^N \Phi ({\bf r}_a)\right)\Psi .
\label{VF2}
\end{equation}
The unitary operator \begin{math} U \end{math} comutes with the mass density, 
\begin{equation}
U^\dagger \rho ({\bf r})U =\rho ({\bf r}),
\label{VF3}
\end{equation}
but not with the velocity, 
\begin{equation}
U^\dagger {\bf v}({\bf r})U ={\bf v}({\bf r})
+{\bf grad}\Phi ({\bf r}).
\label{VF4}
\end{equation}
Eq.(\ref{VF4}) follows directly from Eqs.(\ref{QFM5}) and (\ref{VF1}). 
Let us now consider the circulation of the velocity around a closed curve 
\begin{math} C \end{math}; i.e. 
\begin{equation}
\Gamma(C)=\oint_C {\bf v}\cdot d{\bf r}.
\label{VF5}
\end{equation}
Eqs.(\ref{VF4}) and (\ref{VF5}) imply the circulation operator 
tranformation law 
\begin{equation}
U^\dagger \Gamma(C)U=\Gamma(C)+
\oint_C {\bf grad}\Phi \cdot d{\bf r}.
\label{VF6}
\end{equation}
The last integral on the right hand side of Eq.(\ref{VF6}) need not 
vanish if the closed curve \begin{math} C \end{math} resides in a multiply 
connected region of nonzero mean fluid density.

For example, suppose a vortex filament with a finite core which is empty of 
fluid molecules. Let \begin{math} C \end{math} completely surround the core. 
From the single valued nature of quantum mechanical wave functions, in 
particular the velocity potential phase in 
\begin{equation}
U\Psi \equiv 
\left[\prod_{a=1}^N e^{iM\Phi ({\bf r}_a)/\hbar}\right]\Psi ,
\label{VF7}
\end{equation}  
one must have the quantized circulation 
\begin{eqnarray}
\oint_C {\bf grad}\Phi \cdot d{\bf r}
=\frac{2\pi \hbar}{M}N(C)\ ,
\nonumber \\
\ \ \ 
\nonumber \\
N(C)=0,\pm 1,\pm 2, \pm 3, \pm 4 \ldots \ \ \ .
\label{VF8}
\end{eqnarray}
The central problem of quantum fluid mechanical turbulence 
concerns the nature of the motions of filaments with  
a quantized circulation in units of 
\begin{math} \kappa = (2\pi \hbar/M) \end{math}.
Let us now turn to a formal Lagrangian 
representation of a field theoretical formulation 
of fluid mechanics.

\section{Clebsch Quantization \label{CRQ}}

Let us at first begin with the {\em classical fluid mechanical  
action principle}\cite{Yourgrau:1979} for adiabatic flows. Later the action will 
be employed to quantize the theory, The results are consistent with 
the Landau quantization approach discussed in the above Sec.\ref{QFM}. 
From a classical fluid mechanical viewpoint, the velocity field 
may be written in the form 
\begin{equation}
{\bf v}={\bf grad}\Phi +\lambda {\bf grad}\mu 
\label{CRQ1}
\end{equation}
wherein the three component velocity fields 
\begin{math} (v_x,v_y,v_z)  \end{math} are replaced by  
three scalar fields \begin{math} \Phi \end{math}, 
\begin{math} \lambda \end{math} and 
\begin{math} \mu \end{math}. The vorticity may then be computed as 
\begin{equation}
{\bf \Omega}={\rm curl}{\bf v}=
{\bf grad}\lambda  \times {\bf grad}\mu . 
\label{CRQ2}
\end{equation}
Vortex lines may be pictured as the intersection between 
two surfaces where one of the surfaces has spatially constant 
\begin{math} \lambda  \end{math} and the other surface has spatially 
constant \begin{math} \mu  \end{math}. In adiabatic flows, one expects 
that the vortex lines move with the fluid. This expectation may be 
implemented by allowing {\em three classical conservation laws}; i.e. 
\begin{eqnarray}
\frac{\partial \rho}{\partial t}+{\rm div}(\rho {\bf v})=0,
\nonumber \\ 
\frac{\partial (\rho \lambda )}{\partial t}
+{\rm div}\big((\rho \lambda ){\bf v}\big)=0,
\nonumber \\ 
\frac{\partial (\rho \mu )}{\partial t}
+{\rm div}\big((\rho \mu ){\bf v}\big)=0.
\label{CRQ3}
\end{eqnarray}
In terms of the total time derivative
\begin{equation}
\frac{d}{dt}=\frac{\partial}{\partial t}
+{\bf v\cdot grad}\ , 
\label{CRQ4}
\end{equation}
we have the conservation laws in the more simple form
\begin{eqnarray}
\frac{d \rho}{dt}=-\rho {\rm div}{\bf v},
\nonumber \\ 
\frac{d \lambda }{dt}=0,
\nonumber \\ 
\frac{d \mu }{dt}=0.
\label{CRQ5}
\end{eqnarray}
Thus, each of the fluid particles and their  
\begin{math} \lambda \end{math} and 
\begin{math} \mu \end{math} surface coordinate 
assignments are conserved. The spatially constant 
surfaces of both \begin{math} \lambda \end{math} and 
\begin{math} \mu \end{math} move with the fluid and so do 
the vortex lines which are the intersection between these 
surfaces. 

Eqs.(\ref{CRQ5}) are an expression of Kelvin's circulation 
theorem\cite{Kelvin:1849} for adiabatic flows. In particular, 
the circulation around a closed boundary curve 
\begin{math} C=\partial \Sigma \end{math} of a surface 
\begin{math} \Sigma \end{math} may be written 
\begin{equation}
\Gamma(C)=\oint_C {\bf v}\cdot d{\bf r}=
\oint_C (d\Phi +\lambda d\mu )
\label{CRQ6}
\end{equation} 
wherein Eq.(\ref{CRQ1}) has been invoked. If 
\begin{math} \Sigma \end{math} lies within a simply connected 
fluid region of non-zero classical fluid density, then Stokes theorem 
implies 
\begin{equation}
\Gamma(C)=\oint_{C=\partial \Sigma} \lambda d\mu 
= \int_\Sigma d\lambda \wedge d\mu . 
\label{CRQ7}
\end{equation} 
If \begin{math} C \end{math}, 
\begin{math} \Sigma \end{math}, \begin{math} \lambda \end{math},
and \begin{math} \mu \end{math} all move with the adiabatic flow, 
then the circulation \begin{math} \Gamma(C) \end{math} is uniform in time 
in accordance with Kelvin's theorem.

Let \begin{math} u(\rho ,s) \end{math} representing the fluid 
energy per unit mass as a function of the mass density 
\begin{math} \rho  \end{math} and the entropy per unit mass 
\begin{math} s \end{math}. The classical fluid action may be written as  
\begin{equation}
A=\int Ldt
=\int \left\{\int {\cal L}d^3{\bf r} \right\}dt, 
\label{CRQ8}
\end{equation} 
with an adiabatic lagrangian density
\begin{equation}
{\cal L}=\rho \left(\frac{1}{2} |{\bf v}|^2- u(\rho ,s)\right)
+{\cal L}^\prime . 
\label{CRQ9}
\end{equation} 
In the above Eq.(\ref{CRQ9}), the first two terms on the right hand side 
represent the usual Galilean invariant 
``kinetic energy minus potential energy'' form while the conservation 
law {\em constraint lagrangian density} may be written
\begin{equation}
{\cal L}^\prime =
\Phi \left(\frac{d\rho}{dt}+\rho {\rm div}{\bf v}\right)  
-\rho \lambda \left(\frac{d\mu }{dt}\right). 
\label{CRQ10}
\end{equation} 
The velocity potential \begin{math} \Phi \end{math} appears as a Lagrange 
multiplier assuring local conservation of mass, \begin{math} \lambda \end{math}   
appears as a Lagrange multiplier assuring local conservation of 
\begin{math} \mu \end{math} and local conservation of 
\begin{math} \lambda \end{math} is the result of the classical variational action 
equations of motion; i.e. \begin{math} \delta A=0 \end{math}. Let us consider this in 
more detail.

Functionally differentiating the action Eq.(\ref{CRQ8}) with respect to the velocity yields 
Eq.(\ref{CRQ1}) in the form,
\begin{equation}
\frac{\delta A}{\delta {\bf v}}=\rho 
\left({\bf v}-{\bf grad}\Phi -\lambda {\bf grad}\mu \right)=0. 
\label{CRQ11}
\end{equation} 
The conservation law Eqs.(\ref{CRQ5}) follow from the variational action 
principle equations 
\begin{eqnarray}
\frac{\delta A}{\delta \Phi}=
\frac{d \rho}{dt}+\rho {\rm div}{\bf v}=0,
\nonumber \\ 
\frac{\delta A}{\delta \mu}=
\lambda \left(\frac{d \rho}{dt}+\rho {\rm div}{\bf v}\right)
+\rho \frac{d \lambda }{dt}=0,
\nonumber \\ 
\frac{\delta A}{\delta \lambda}=-\rho \frac{d \mu }{dt}=0.
\label{CRQ12}
\end{eqnarray}
Still viewing adiabatic fluid mechanics as a classical field theory, one may 
compute the conjugate momenta to the fields of interest employing the general 
field theoretical relation 
\begin{math} \Pi_\varphi =(\delta A/\delta \dot{\varphi}) \end{math}  
where 
\begin{math} \dot{\varphi}\equiv (\partial \varphi /\partial t )\end{math}. 
In particular, two conjugate field-momenta may be computed via 
\begin{eqnarray}
\Pi_\rho =\frac{\delta A}{\delta \dot{\rho }}=
\frac{\partial {\cal L}}{\partial \dot{\rho }}=\Phi ,
\nonumber \\ 
\Pi_\mu =\frac{\delta A}{\delta \dot{\mu }}=
\frac{\partial {\cal L}}{\partial \dot{\mu }}=-\rho \lambda . 
\label{CRQ13}
\end{eqnarray}
The energy of a classical fluid flow in the lagrangian formalism is 
\begin{equation}
{\rm E}=\int \left\{\dot{\rho}\frac{\partial {\cal L}}{\partial \dot{\rho }} 
+\dot{\mu }\frac{\partial {\cal L}}{\partial \dot{\mu }}-{\cal L}\right\}d^3{\bf r},
\label{CRQ14}
\end{equation}
which after parts integration yields 
\begin{equation}
{\rm E} =\int {\cal E}d^3{\bf r}
\ \ {\rm wherein}\ \ {\cal E}
=\frac{1}{2}\rho |{\bf v}|^2+\rho  u(\rho,s).
\label{CRQ15}
\end{equation}
The quantized fluid mechanical model may now be formulated as follows:
\medskip 
\par \noindent
(i) There are four fields of interest \begin{math} \rho \end{math}, 
\begin{math} \Phi \end{math}, \begin{math} \lambda \end{math}
and \begin{math} \mu \end{math}. Either field of the pair 
\begin{math} (\rho ,\Phi ) \end{math} commutes with either field 
of the pair \begin{math} (\mu ,\lambda ) \end{math}.
From Eqs.(\ref{CRQ13}), we have the two conjugate field commutation 
relations pairs 
\begin{eqnarray}
\frac{i}{\hbar }\left[\Phi ({\bf r}),\rho ({\bf r}^\prime )\right]
=\delta ({\bf r}-{\bf r}^\prime ), 
\nonumber \\ 
\frac{i}{\hbar }\left[\mu ({\bf r}),\lambda ({\bf r}^\prime )\right]
=\frac{\delta ({\bf r}-{\bf r}^\prime )}{\rho (\bf r)}.
\label{CRQ16}
\end{eqnarray}
The fluid velocity field operator is computed via Eq.(\ref{CRQ1}), employing   
the operator ordering
\begin{equation}
{\bf v}({\bf r})={\bf grad}\Phi ({\bf r})+\frac{1}{2}
\{\lambda ({\bf r}),{\bf grad}\mu (\bf r)\}
\label{CRQ17}
\end{equation}
where \begin{math} \{A,B\}\equiv AB+BA  \end{math} and
with implicit point splitting for field multiplications at the same 
spatial point. From Eqs.(\ref{CRQ16}) and (\ref{CRQ17}), one may recover 
the Landau equal time commutation Eqs.(\ref{QFM5}), (\ref{QFM7}) and (\ref{QFM8}).
\medskip 
\par \noindent
(ii) The Hamiltonian follows from Eq.(\ref{CRQ15}) to be 
\begin{equation}
H=\int {\cal H}d^3{\bf r}
\ \ {\rm wherein}\ \ {\cal H}
=\frac{1}{2}{\bf v}\cdot ({\bf 1} \rho )\cdot {\bf v}+\rho  u(\rho,s).
\label{CRQ18}
\end{equation}
Note the operator ordering between the non-commuting field operators 
\begin{math} \rho \end{math} and \begin{math} {\bf v} \end{math}.
\medskip 
\par \noindent
(iii) The fluid temperature \begin{math} T \end{math} and pressure 
\begin{math} P \end{math} may be computed from the thermodynamic law 
\begin{equation}
du =Tds+\frac{P}{\rho^2}d\rho .
\label{CRQ19}
\end{equation} 
In terms of the enthalpy per unit mass, 
\begin{equation}
w=\left[\frac{\partial (\rho u)}{\partial \rho }\right]_s
=u +\frac{P}{\rho }, 
\label{CRQ20}
\end{equation} 
we have the thermodynamic law 
\begin{equation}
dw=Tds+\frac{1}{\rho }dP.
\label{CRQ21}
\end{equation}
For adiabatic flows \begin{math} ds=0  \end{math}; i.e. 
\begin{equation}
\rho\ {\bf grad}w={\bf grad}P
\ \ \ \ \ {\rm (adiabatic)}.
\label{CRQ22}
\end{equation}
\medskip 
\par \noindent
(iv) The equation of motion for the mass density may be derived 
employing 
\begin{equation}
\dot{\rho }({\bf r})
=\frac{i}{\hbar }\left[H,\rho ({\bf r})\right]
=-{\rm div}{\bf J}({\bf r})
\label{CRQ23}
\end{equation} 
wherein the current density operator 
\begin{math} {\bf J}  \end{math}
is defined in Eq.(\ref{QFM4}). The equation of motion for the 
velocity field, 
\begin{equation}
\dot{\bf v}({\bf r})
=\frac{i}{\hbar }\left[H,{\bf v}({\bf r})\right],
\label{CRQ24}
\end{equation} 
has the quantum operator form 
\begin{eqnarray}
\dot{\bf v}({\bf r})+
{\bf grad}\big(\frac{1}{2}{\bf v}({\bf r})\cdot {\bf v}({\bf r})
+w(\rho ({\bf r}),s)\big)
\nonumber \\ 
+\frac{1}{2}\big({\bf \Omega }({\bf r})\times {\bf v}({\bf r})
-{\bf v}({\bf r})\times {\bf \Omega }({\bf r})\big)  =0,
\label{CRQ25}
\end{eqnarray}
with the vorticity \begin{math} {\bf \Omega}={\rm curl}{\bf v} \end{math}.
Eq.(\ref{CRQ25}) is simply the quantum field theoretical version of 
Newton's law, 
\begin{equation}
\rho \frac{d{\bf v}}{dt}=-{\bf grad}P
\ \ \ \ \ {\rm (classical)},
\label{CRQ26}
\end{equation}
in thinly disguised form. In Eq.(\ref{CRQ26}) we have invoked 
Eqs.(\ref{CRQ4}) and (\ref{CRQ22}). Finally, the equation of motion for 
vorticity follows by taking the curl of Eq.(\ref{CRQ25}); i.e.  
\begin{eqnarray}
\dot{\bf \Omega }({\bf r})=
\frac{i}{\hbar }\left[H,{\bf \Omega}({\bf r})\right],
\nonumber \\ 
\dot{\bf \Omega }({\bf r})+\frac{1}{2}
{\rm curl}\big({\bf \Omega}({\bf r})\times {\bf v}({\bf r})
-{\bf v}({\bf r})\times {\bf \Omega}({\bf r})\big)=0.
\label{CRQ27}
\end{eqnarray}

This completes the local lagrangian formulation of quantum fluid mechanics 
for adiabatic flows. The quantum field theory has been well defined 
modulo the usual list of suspect mathematical rigor problems endemic to 
strongly non-linear quantum field theories describing an infinite number 
of degrees of freedom.

\section{Kolmogorov Correlations\label{KC}}

Consider the velocity fields of fully developed turbulence in a frame of 
reference in which the mean drift velocity are zero. The statistical correlations 
of the velocity field for {\em isotropic incompressible turbulence} 
(\begin{math} div{\bf v}=0 \end{math}) may be described by 
\begin{eqnarray}
{\sf G}({\bf r}-{\bf r}^\prime )=\frac{1}{2}
\big<\{{\bf v}({\bf r}),{\bf v}({\bf r}^\prime )\}\big>
\nonumber \\ 
{\sf G}({\bf r}-{\bf r}^\prime )
=\int {\sf S}({\bf k})e^{i{\bf k}\cdot ({\bf r}-{\bf r}^\prime )}
\left[\frac{d^3{\bf k}}{(2\pi )^3}\right],
\nonumber \\ 
{\sf S}({\bf k})=\left({\sf 1}-\frac{\bf kk}{k^2}\right)S(k),
\label{KC1a}
\end{eqnarray}
wherein \begin{math} \{a,b\}\equiv (ab+ba) \end{math} and 
\begin{eqnarray}
G(|{\bf r}-{\bf r}^\prime |)=tr{\sf G}({\bf r}-{\bf r}^\prime ),
\nonumber \\ 
G(|{\bf r}-{\bf r}^\prime |)=\frac{1}{2}
\big<{\bf v}({\bf r})\cdot {\bf v}({\bf r}^\prime )+
{\bf v}({\bf r}^\prime )\cdot {\bf v}({\bf r})\big>,
\nonumber \\ 
G(r)=\frac{1}{\pi^2 }\int_0^\infty k^2 
S(k)\left[\frac{\sin(kr)}{kr}\right]dk.
\label{KC1b}
\end{eqnarray}
The Komlmogorov energy spectral distribution \begin{math} {\cal E}(k)dk \end{math} 
is defined as the portion of the kinetic energy per unit mass in the wavenumber 
width \begin{math} dk \end{math}; i.e. Eqs.(\ref{KC1a}) and (\ref{KC1b}) imply 
\begin{eqnarray}
{\cal E}(k)=\left[\frac{k^2 S(k)}{2\pi^2}\right],
\nonumber \\ 
G(r)=2\int_0^\infty {\cal E}(k)\left[\frac{\sin(kr)}{kr}\right]dk.
\label{KC2}
\end{eqnarray}
The mean squared velocity of the isotropic turbulence is thereby 
\begin{equation}
\frac{1}{2}\left<\left|{\bf v}\right|^2 \right>=
\int_0^\infty {\cal E}(k)dk.
\label{KC3}
\end{equation}
Vorticity correlations may be discussed in a similar manner  
\begin{eqnarray}
{\sf F}({\bf r}-{\bf r}^\prime )=\frac{1}{2}
\big<\{{\bf \Omega}({\bf r}),{\bf \Omega }({\bf r}^\prime )\}\big>
\nonumber \\ 
{\sf F}({\bf r}-{\bf r}^\prime )
=\int {\sf S}_{\bf \Omega }({\bf k})e^{i{\bf k}\cdot ({\bf r}-{\bf r}^\prime )}
\left[\frac{d^3{\bf k}}{(2\pi )^3}\right],
\nonumber \\ 
{\sf S}_{\bf \Omega }({\bf k})=\left({\sf 1}-\frac{\bf kk}{k^2}\right)S_{\bf \Omega}(k),
\nonumber \\ 
S_{\bf \Omega}(k)=k^2S(k).
\label{KC1av}
\end{eqnarray}
If we allow  
\begin{eqnarray}
F(|{\bf r}-{\bf r}^\prime |)=tr{\sf F}({\bf r}-{\bf r}^\prime ),
\nonumber \\ 
F(|{\bf r}-{\bf r}^\prime |)=\frac{1}{2}
\big<{\bf \Omega }({\bf r})\cdot {\bf \Omega }({\bf r}^\prime )+
{\bf \Omega }({\bf r}^\prime )\cdot {\bf \Omega }({\bf r})\big>,
\nonumber \\ 
F(r)=\frac{1}{\pi^2 }\int_0^\infty k^2 
S_{\bf \Omega}(k)\left[\frac{\sin(kr)}{kr}\right]dk,
\label{KC1bv}
\end{eqnarray}
then the spectral theorems hold true;
\begin{eqnarray}
rF(r)=2\int_0^\infty k{\cal E}(k)\sin(kr)dk, 
\nonumber \\ 
k{\cal E}(k)=\frac{1}{\pi }\int_0^\infty rF)r)\sin(kr)dr.
\label{KC1cv} 
\end{eqnarray}
The mean squared turbulent vorticity may also be computed as 
\begin{equation}
\omega^2\equiv \left<\left|{\bf \Omega }\right|^2 \right>
=2\int_0^\infty k^2 {\cal E}(k)dk.
\label{KC4}
\end{equation} 

The {\em classical} viscous heating rate per unit mass 
from the turbulent mass eddy currents may be taken to be 
\begin{equation}
\epsilon = \frac{\eta \omega^2}{\rho }=\nu \omega^2 , 
\label{KC5}
\end{equation}
wherein \begin{math} \eta  \end{math} is the fluid viscosity. 
Kolmogorov\cite{Kolmogorov:1941c} assumed that the energy spectral function depends only 
on the wave number \begin{math} k \end{math} and the dissipation 
per unit mass \begin{math} \epsilon \end{math}, i.e. 
\begin{equation}
{\cal E}={\cal F}(k,\epsilon),
\label{KC6}
\end{equation} 
and then considered the physical dimensions of the quantities involved.
These dimensions (in cgs units) are 
\begin{eqnarray}
{\rm dim}[{\cal E}] = \frac{\rm cm^3}{\rm sec^2}\ ,
\nonumber \\ 
{\rm dim}[\epsilon ] = \frac{\rm cm^2}{\rm sec^3}\ ,
\nonumber \\ 
{\rm dim}[k] = \frac{1}{\rm cm}\ .
\label{KC7}
\end{eqnarray}
The only functional form in Eq.(\ref{KC6}) with the correct physical 
dimensions of Eq.(\ref{KC7}) is given by 
\begin{equation}
{\cal E}(k)=C\left[\frac{\epsilon^{2/3}}{k^{5/3}}\right],
\label{KC8}
\end{equation}
where \begin{math} C  \end{math} is a dimensionless constant presumably 
of order unity\cite{Sreenivasan:1995}. The correlation between turbulent 
velocities at two different points may be defined as 
\begin{eqnarray}
g({\bf r}-{\bf r}^\prime)=
\left<|{\bf v}({\bf r})-{\bf v}({\bf r}^\prime )|^2\right>,
\nonumber \\ 
g(r)=2(G(0)-G(r)),
\nonumber \\ 
g(r)=4\int_0^\infty 
{\cal E}(k)\left[1-\frac{\sin(kr)}{kr}\right]dk,
\label{KC9}
\end{eqnarray}
wherein Eq.(\ref{KC2}) has been invoked. 
Employing Eqs.(\ref{KC8}) and (\ref{KC9}), one finds 
\begin{eqnarray}
C^\prime =4C\int_0^\infty \frac{1}{\beta ^{5/3}}
\left[1-\frac{\sin \beta }{\beta } \right]d\beta ,
\nonumber \\ 
C^\prime =\frac{9}{5}\Gamma\left(\frac{1}{3}\right)C,
\nonumber \\ 
g(r)=C^\prime (\epsilon r)^{2/3}, 
\label{KC10}
\end{eqnarray}
wherein the gamma function, 
\begin{equation}
\Gamma (z)=\int_0^\infty s^z e^{-s}\frac{ds}{s}\ , 
\label{KC11}
\end{equation}
has been invoked. Under the assumption that 
\begin{math} g={\cal G}(r,\epsilon ) \end{math} 
alone, the proportionality in Eq.(\ref{KC10}), i.e. 
\begin{math} g(r)\propto (\epsilon r)^{2/3} \end{math} 
follows from the analysis of physical dimensions. 
From Eqs.(\ref{KC1cv}) and (\ref{KC8}) follows the Kolmogorov 
correlation scaling function for vorticity
\begin{eqnarray}
\tilde{C}=\Gamma\left(\frac{1}{3}\right)C=\frac{5}{9}C^\prime ,
\nonumber \\ 
F(r)=\tilde{C}\frac{\epsilon ^{2/3}}{r^{4/3}}\ . 
\label{KC12}
\end{eqnarray}
Let us now see how the Kolmogorov scaling Eq.(\ref{KC12}) may arise from 
quantum vortex filaments.

\section{Quantum Vortex Filaments \label{FDT}}

The scaling laws in Eqs.(\ref{KC8}) and (\ref{KC10}) appear to be valid 
for fully developed turbulence in superfluid Helium\cite{Maurer:1998, 
Stalp:1999, Roche:2007, Bradley:2008} as 
well as in liquids normally regarded as classical in nature. Let us see 
why this may be so from the viewpoint of quantum fluid mechanical vortex 
filaments.
From Eq.(\ref{VF8}), it follows that the circulation around a single 
quantum fluid vortex obeys 
\begin{equation}
\oint {\bf v}\cdot d{\bf r}=\kappa \equiv \frac{2\pi \hbar }{M}\ .
\label{FDT1}
\end{equation} 
The magnitude of fluid vorticity is related to the number of quantum vortex 
per unit area \begin{math} {\cal A}^{-1} \end{math} via 
\begin{math} \Omega =(\kappa /{\cal A}) \end{math}. If we examine just one quantum 
vortex filament and let \begin{math} {\bf T} \end{math} be the unit tangent vector
to the quantized filament, then 
\begin{equation}
{\bf \Omega }=\frac{\kappa }{\cal A}{\bf T}.
\label{FDT2}
\end{equation}
Following this single quantum vortex filament, let \begin{math} s \end{math}
denote the arc length alongh the filament path so that 
\begin{equation}
{\bf T}(s)=\frac{d {\bf r}(s)}{ds}.
\label{FDT3}
\end{equation}
Along a single quantum vortex filament, let us consider the correlation function 
\begin{equation}
H(s_1-s_2)=\frac{1}{2}\left<{\bf T}(s_1)\cdot {\bf T}(s_2)
+{\bf T}(s_2)\cdot {\bf T}(s_1)\right>;
\label{FDT4}
\end{equation}
From Eqs.(\ref{FDT3}) and (\ref{FDT4}) it follows that the mean square distance between 
two points on the fillament separated by a filament length \begin{math} L \end{math} 
is given by 
\begin{equation}
R^2=\left<\left|{\bf r}_f-{\bf r}_i\right|^2\right>=
\int_L  \int_L H(s_1-s_2)ds_1ds_2.
\label{FDT5}
\end{equation}
If the filament path is of fractal dimension \begin{math} d \end{math}, 
then the mean square radius varies with \begin{math} L \end{math} according to 
\begin{equation}
\frac{L}{\xi }=\left(\frac{R}{\xi}\right)^d
\ \ \ \Rightarrow \ \ \ R^2=\xi^2 \left(\frac{L}{\xi}\right)^{2/d},
\label{FDT6}
\end{equation}
wherein \begin{math} \xi \end{math} is a length scale along the filament core  
required to achieve an appreciable bending. Evidently such a length should be 
large on the scale of the diameter of the filament core; i.e. 
\begin{math} \xi \gg \sqrt{\cal A}  \end{math}. In order that Eqs.(\ref{FDT5}) 
and (\ref{FDT6}) hold true, one may write  
\begin{equation}
H(L)=\frac{1}{2}\left[\frac{\xi }{L}\right]^{2(d-1)/d}
\label{FDT7}
\end{equation}
Employing Eqs.(\ref{KC1bv}), (\ref{FDT2}), (\ref{FDT4}) and (\ref{FDT7}), 
one finds for fractal dimension \begin{math} d \end{math}  
\begin{equation}
F_d(R)=\left(\frac{\kappa }{\cal A}\right)^2 H(L)
=\frac{1}{2}\left(\frac{\kappa }{\cal A}\right)^2 \left[\frac{\xi }{L}\right]^{2(d-1)/d},
\label{FDT8}
\end{equation}
which now reads 
\begin{equation}
F_d(R)=\frac{1}{2}\left(\frac{\kappa }{\cal A}\right)^2 
\left(\frac{\xi }{R}\right)^{2(d-1)}.
\label{FDT9}
\end{equation}
in virtue of Eqs.(\ref{FDT6}) and (\ref{FDT8}). Eq.(\ref{FDT9}), describing vorticity 
correlations by employing quantized vortex filaments of fractal dimension 
\begin{math} d \end{math}, is the central result of this section.  

If the fractal dimension of the quanized vortex is that of a self avoiding random walk, 
as in Eq.(\ref{Intro6}), then \begin{math} F(r) \end{math} is given by 
\begin{equation}
F(r)=\frac{1}{2}\left(\frac{\kappa }{\cal A}\right)^2 
\left(\frac{\xi }{r}\right)^{4/3} 
\label{FDT10}
\end{equation}
which exhibits the Kolmogorov scaling as in Eq.(\ref{KC12}).

\section{Conclusion \label{CONC}}

A quantized theory of fluid mechanics was reviewed and formulated 
as a Lagrangian quantum field theory.  Quantized 
vortex filaments were discussed.  A theory of the quantized fluid
mechanics was written in the Clebsch representation.  Finally, a
model of quantized vortices was given wherein the Kolmogorov 
scaling law was derived.

Recent studies of turbulence in superfluid Helium indicate that  
turbulence in quantum fluids obeys a Kolmogorov scaling law. 
In this paper, it is suggested that turbulence in all fluids is due 
to quantum fluid mechanical effects.  There have been found to
be many differences\cite{Tsubota:2008, Finne:2003, Paoletti:2008}
between turbulence in superfluid He and 
other fluids, yet more striking, are the similarities\cite{Vinen:2000}. 
It has been found experimentally that superfluid \begin{math} ^4He_2 \end{math}
and more recently liquid helium \begin{math} ^3He-B \end{math} exhibit
the Kolmogorov scaling law found in other fluids, an unexpected result.  
The fact that both ``classical'' and quantum fluids manifest the same scaling laws,
suggests that all fluids need to be treated using a quantum mechanical
formalism when turbulence takes place. 

Turbulent flows occur
when the Reynold's number becomes large, i.e., \begin{math}{\cal R}
>>1\end{math}.  In this regime, the kinematic viscosity is small.
Although turbulent flows are considered highly dissipative\cite{Tsinober:2001},
they last a long time, even after an external energy source ceases to exist.  
A vanishing kinematic viscosity implies that the ratio ${\cal R}_F/{\cal R}=\nu/\kappa$
is relatively small.    
The implication is that there are few quantum vortices in a classical bundle. 
The quantum vortices themselves obey the Kolmogorov 5/3 scaling law.


\begin{thebibliography}{07}

\bibitem{LandauLifshitz:1987}
L.D. Landau and E.M. Lifshitz, ``Fluid Mechanics'',
Chapter III, Pergamon Press, Oxford (1987).  

\bibitem{McComb:1990}
W.D. McComb, ``The Physics of Fluid Turbulence'',
Oxford University Press, Oxford (1990).

\bibitem{Feynman:1955}
R.P. Feynman, {\it Prog. Low Temp. Phys.} {\bf 1}, 
17 (1955).

\bibitem{Widom:1990}
A.~Widom and Y.~N.~Srivastava, {\it Modern Physics Letters B} {\bf 4}, 1 (1990).

\bibitem{Landau:1941}
L.D. Landau, {\it J. Phys. USSR} {\bf 5}, 77 (1941). 

\bibitem{Clebsch:1859}
A. Clebsch, {\it J. Reine Angew. Math} {\bf 56}, 1 (1859).

\bibitem{Kolmogorov:1941a}
A.N. Kolmogorov, {\it Dok. Akad. Nauk. SSSR} {\bf 30}, 299 (1941).   

\bibitem{Kolmogorov:1941b}
A.N. Kolmogorov, {\it Dok. Akad. Nauk. SSSR} {\bf 31}, 538 (1941).

\bibitem{Kolmogorov:1941c}
A.N. Kolmogorov, {\it Dok. Akad. Nauk. SSSR} {\bf 32}, 16 (1941).

\bibitem{Yourgrau:1979}
W.~Yourgrau and S.~Mandelstam, ``Variational Principles in Dynamics and Quantum Theory'', {\it Dover Publication, Inc.}, p. 147 (1979).

\bibitem{Kelvin:1849}
W. Thomson, ``On the Vis-Viva of a Liquid in Motion'',
{\it Cambridge and Dublin Mathematics Journal}, (1849). 

\bibitem{Sreenivasan:1995}
K.~R.~Sreenivasan, {Phys. Fluids} {\bf 7}, 2778 (1995). 

\bibitem{Maurer:1998}
J.~Maurer and P.~Tabeling, {\it Europhys Lett} {\bf 43}, 29 (1998).

\bibitem{Stalp:1999}
S.~R.~Stalp, L.~Skrbek, and Russell J.~Donnelly, {\it Physical Review Letters} {\bf 82}, 4831 (1999).

\bibitem{Roche:2007}
P.~E.~Roche, et al., {\it EPL} {\bf 77} 66002 (2007).

\bibitem{Bradley:2008}
D.~I.~Bradley, et al., {\it Physical Review Letters} {\bf 101}, 065302 (2008).

\bibitem{Chorin:1990}
A.~J.~Chorin, {\it Commun. Math. Phys.} {\bf 132}, p. 519 (1990).

\bibitem{Tsubota:2008}
Makoto~Tsubota, {\it J. Phys. Soc. Jpn} {\bf 77}, 1 (2008).

\bibitem{Finne:2003}
A.~Finne, et al., {\it Nature} {\bf 424}, 1022 (2003).

\bibitem{Paoletti:2008}
M.~S.~Paoletti, et al., {\it Physical Review Letters}, {\bf 101}, 154501 (2008).

\bibitem{Vinen:2000}
W.~F.~Vinen, {\it Physical Review B} {\bf 61}, 1410 (2000).

\bibitem{Tsinober:2001}
A.~Tsinober, ``An Informal Introduction to Turbulence'', p. 17, Kluwer Academic Publishers (2001).


\end{thebibliography}
\end{document}